\documentclass[prd,twocolumn,a4paper,superscriptaddress,showpacs]{revtex4}

\usepackage{setspace}
\usepackage{amsmath,amssymb,amsfonts,amsthm}
\usepackage{graphicx}
\usepackage{enumerate}

\begin{document}

\title{Black hole state counting in Loop Quantum Gravity: A number theoretical approach}

\author{Iv\'an  \surname{Agull\'o}}
\email[]{Ivan.Agullo@uv.es} \affiliation{Departamento de F\'{\i}sica
Te\'orica and IFIC, Centro Mixto Universidad de Valencia-CSIC.
Facultad de F\'{\i}sica, Universidad de Valencia, Burjassot-46100,
Valencia, Spain}

\author{J. Fernando \surname{Barbero G.}}
\email[]{fbarbero@iem.cfmac.csic.es} \affiliation{Instituto de
Estructura de la Materia, CSIC, Serrano 123, 28006 Madrid, Spain}

\author{Jacobo  \surname{D\'{\i}az-Polo}}
\email[]{Jacobo.Diaz@uv.es} \affiliation{Departamento de
Astronom\'{\i}a y Astrof\'{\i}sica, Universidad de Valencia,
Burjassot-46100, Valencia, Spain}

\author{Enrique  \surname{Fern\'andez-Borja}}
\email[]{Enrique.Fernandez@uv.es} \affiliation{Departamento de
F\'{\i}sica Te\'orica and IFIC, Centro Mixto Universidad de
Valencia-CSIC. Facultad de F\'{\i}sica, Universidad de Valencia,
Burjassot-46100, Valencia, Spain} \affiliation{Departamento de
Astronom\'{\i}a y Astrof\'{\i}sica, Universidad de Valencia,
Burjassot-46100, Valencia, Spain}

\author{Eduardo J. \surname{S. Villase\~nor}}
\email[]{ejsanche@math.uc3m.es} \affiliation{Instituto Gregorio Mill\'an, Grupo de Modelizaci\'on
y Simulaci\'on Num\'erica, Universidad Carlos III de Madrid, Avda.
de la Universidad 30, 28911 Legan\'es, Spain} \affiliation{Instituto
de Estructura de la Materia, CSIC, Serrano 123, 28006 Madrid, Spain}

\date{March 12, 2008}

\begin{abstract}
We give an efficient method, combining number theoretic and combinatorial ideas, to \textit{exactly} compute black hole entropy in the framework of Loop Quantum Gravity. Along the way we
provide a complete characterization of the relevant sector of the
spectrum of the area operator, including degeneracies, and \textit{explicitly} determine the
number of solutions to the projection constraint. We use a computer implementation of the
proposed algorithm to confirm and extend previous results on the
detailed structure of the black hole degeneracy spectrum.
\end{abstract}


\maketitle

Any proposed quantum theory of gravity must account for the states
responsible for black hole entropy. Within Loop Quantum Gravity
(LQG) entropy can be studied by using the Isolated Horizon
framework, \cite{abk}. The counting of states is reduced in this
setting to a well defined combinatorial problem. It gives rise, in
the asymptotic limit, to the semiclassical Bekenstein-Hawking
formula \cite{DL,GM} corrected to the next relevant order by a term
logarithmic in the area. A computer assisted study has been carried
out for small black holes up to two hundreds of Planck areas $\ell_P^2$
\cite{val1}. This has unearthed a very interesting behavior in
their degeneracy spectrum, namely, an equidistant ``band structure''
with important physical consequences. The most relevant of them is
the effective quantization of black hole entropy \cite{val1}. A qualitative
understanding of the origin of this behavior has been obtained in \cite{val2}.
However, no detailed theoretical description of this phenomenon has been
available to date owing to the incomplete characterization of the
area spectrum, on one hand, and the lack of exact manageable
solutions for some combinatorial problems (involving the so called
projection constraint), on the other.

In this letter we present a satisfactory solution to both types of
difficulties, giving a precise characterization of the area spectrum
by relying on number-theoretic methods, and addressing the
combinatorial problems related to the projection constraint. We do it for the original counting of states proposed in \cite{abk} and carried out in \cite{DL}, and also for the one described in \cite{GM}. The method that we discuss in the following will
allow us to have a full understanding of the different factors that
come into play to reproduce the features previously observed in the
black hole degeneracy spectrum. In addition, it can be efficiently
used to perform \textit{exact} entropy computations --extensible to
large areas-- that improve and confirm the results obtained by brute
force methods in \cite{val1}.

We start by characterizing the area eigenvalues and their
degeneracies. In LQG the black hole area is given by an eigenvalue
$A$ of the area operator
\begin{equation}
A=8\pi\gamma
\ell_P^2\sum_{I=1}^N\sqrt{j_I(j_I+1)}\,,\label{area_eigenvalue}
\end{equation}
where $\gamma$ denotes the Immirzi parameter. Notice that these do not give the full area spectrum but, for the case of isolated horizons relevant here we only need equation (\ref{area_eigenvalue}). The labels $j_I$ are half-integers, $j_I\in\mathbb{N}/2$, associated to the edges of
a given spin network state. They pierce the horizon at a finite set
of $N$ distinguishable points called punctures \cite{abk}. Horizon quantum states
are further characterized by an additional label $m_I$. In the case
where we have spherical symmetry a \textit{projection constraint}
\begin{equation}
\sum_{I=1}^Nm_I=0\label{proy_const}
\end{equation}
must be satisfied by the $m_I$. There are two inequivalent proposals
in the literature to account for the relevant microscopic
configurations \cite{DL,GM}. When taken as a purely combinatorial
problem, they differ in the range of the label $m_I$. In the standard (DLM) counting performed in \cite{DL} one takes $m_I\in\{- j_I,j_I\}$, whereas the counting proposed in \cite{GM} (that we will refer to as the GM counting) assumes that $m_I$ can take all the allowed values for a spin component
$m_I\in\{-j_I,-j_I+1,\ldots,j_I-1,j_I\}$.

A first problem that we address is the characterization of the
numbers belonging to the spectrum of the area operator restricted to the vector subspace spanned by spin network states having no vertices nor edges lying on the black hole horizon. In the following when we talk about the area spectrum we refer, in fact, to this restriction.
The first question that we want to consider is: Given $A\in{\mathbb{R}}$, when does it belong to the spectrum of the area? In order to simplify the algebra  and work with integer numbers we
will write $j_I=k_I/2$ in the following, so that the area
eigenvalues become
$$
A=\sum_{I=1}^N\sqrt{(k_I+1)^2-1}=\sum_{k=1}^{k_{\mathrm{max}}}n_k\sqrt{(k+1)^2-1}.
$$
Here we have chosen units such that $4\pi\gamma \ell_P^2=1$, and
the $n_k$ (satisfying $n_1+\cdots n_{k_{\mathrm{max}}}=N$) denote
the number of punctures corresponding to edges carrying spin $k/2$.
An elementary but useful comment is that we can always write
$\sqrt{(k+1)^2-1}$ as the product of an integer and the square root
of a square-free positive integer number (SRSFN) by using its prime
factor decomposition. Hence, with our choice of
units, only integer linear combinations of SRSFN's can appear in the
area spectrum. The questions now are: First, given such a linear
combination,  when does it correspond to an eigenvalue of the area
operator? If the answer is in the affirmative, what are the
permissible choices of $k$ and $n_k$ compatible with this value for
the area?\\
In the following we will take advantage of the fact that SRSFN's are
linearly independent over the rational numbers (and, hence, over the
integers) i.e. $q_1\sqrt{p_1}+\cdots+q_r\sqrt{p_r}=0$, with
$q_i\in\mathbb{Q}$ and $p_i$ different square-free integers, implies
that $q_i=0$ for every $i=1,\ldots,r$. This can be easily checked
for concrete choices of the $p_i$ and can be proved in general (see
for instance \cite{newman}). We can answer the two questions posed
above in the following way. Given an \textit{integer} linear
combination of SRSFN's $\sum_{i=1}^r q_i \sqrt{p_i}$, where
$q_i\in\mathbb{N}$, we need to determine the values of the $k$ and
$n_k$, if any, that solve the equation
\begin{equation}
\label{ecuacion_fund}
\sum_{k=1}^{k_{\mathrm{max}}}n_k\sqrt{(k+1)^2-1}=\sum_{i=1}^r q_i
\sqrt{p_i}\,.
\end{equation}
Each $\sqrt{(k+1)^2-1}$ can be written as an integer times a SRSFN
so the left hand side of (\ref{ecuacion_fund}) will also be a
linear combination of SRSFN with coefficients given by integer
linear combinations of the unknowns $n_k$. As a preliminary step,
let us find out  --for a given square-free positive integer $p_i$--
the values of $k$ satisfying
\begin{equation}
\label{pell} \sqrt{(k+1)^2-1}=y \sqrt{p_i}\,,
\end{equation}
for some positive integer $y$. This is equivalent  to
solving the Pell equation $x^2-p_iy^2=1$ where the unknowns are
$x:=k+1$ and $y$. Equation (\ref{pell}) admits an infinite number of
solutions $(k_m^i,y_m^i)$,  where $m\in\mathbb{N}$ (see, for instance, \cite{Burton}).
These can be
obtained from the fundamental one $(k_1^i,y_1^i)$ corresponding to
the minimum, non-trivial, value of both $k_m^i$ and $y_m^i$. They
are given by the formula
$$
k_m^i+1+y_m^i\sqrt{p_i}=(k_1^i+1+y_1^i\sqrt{p_i})^m.
$$
The fundamental solution can be obtained by using continued
fractions \cite{Burton}. Tables of the fundamental solution for the
smallest $p_i$ can be found in standard references on number theory.
As we can see both $k_m^i$ and $y_m^i$ grow
exponentially in $m$. By solving the Pell equation for all the different
$p_i$ we can rewrite (\ref{ecuacion_fund}) as
$$
\sum_{i=1}^r\sum_{m=1}^\infty n_{k_m^i} y_m^i\sqrt{p_i}=\sum_{i=1}^r
q_i \sqrt{p_i}.
$$
Using the linear independence of the $\sqrt{p_i}$, the previous
equation can be split into $r$ different equations of the type
\begin{equation}
\sum_{m=1}^\infty y_m^i n_{k_m^i}=q_i,\quad i=1,\ldots, r.
\label{diof}
\end{equation}
Several comments are in order now. First, these are diophantine
linear equations in the unknowns $n_{k_m^i}$ with the solutions
restricted to take non-negative values. They can be solved by
standard algorithms (for example the Fr\"obenius method or
techniques based on the use of Smith canonical forms). These are
implemented in commercial symbolic computing packages. Second,
although we have extended the sum in (\ref{diof}) to infinity it is
actually finite because the $y_m^i$ grow with $m$ without bound.
Third, for different values of $i$ the equations (\ref{diof}) are
written in terms of disjoint sets of unknowns. This means that they
can be solved independently of each other --a very convenient fact
when performing actual computations. Indeed, if
$(k^{i_1}_{m_1},y^{i_1}_{m_1})$ and $(k^{i_2}_{m_2},y^{i_2}_{m_2})$
are solutions to the Pell equations associated to different
square-free integers $p_{i_1}$ and $p_{i_2}$, then $k^{i_1}_{m_1}$
and $k^{i_2}_{m_2}$ must be different. This can be easily proved by
\textit{reductio ad absurdum}.

It may happen that some of the equations in (\ref{diof}) admit no
solutions. In this case $\sum_{i=1}^r q_i \sqrt{p_i}$ does not
belong to the relevant part of the area spectrum. On the other hand, if
these equations do admit solutions, the $\sum_{i=1}^r q_i
\sqrt{p_i}$ belong to the spectrum of the area operator, the numbers
$k_m^i$ tell us the spins involved, and the  $n_{k_m^i}$ count the
number of times that the edges labeled by the spin $k_m^i/2$ pierce
the horizon. A set of pairs $\{(k_m^i,n_{k_m^i})\}$ obtained from the
solutions to equations (\ref{ecuacion_fund}), (\ref{pell}), and (\ref{diof}) will define what we call a \textit{spin configuration}. The number of different quantum states
associated to each of these is given by two
degeneracy factors, namely, the one coming from reorderings of  the
$k_I$-labels over the distinguishable punctures (r-degeneracy) and the
other originating in all the different choices of $m_I$-labels
satisfying (2), (m-degeneracy). The combinatorial factors associated
to the r-degeneracy are straightforward to obtain and appear in the
relevant literature.

Let us consider then the m-degeneracy. The problem that we have to
solve is: Given a set of (possibly equal) spin labels $j_I$,
$I=1,\ldots,N$,  what are the different choices for the allowed
$m_I$ such that (\ref{proy_const}) is satisfied? Notice that an
obvious necessary condition for the existence of solutions is that
$\sum_{I=1}^Nj_I\in \mathbb{N}$.

In the standard DLM approach the number of different solutions for the
projection constraint can be found by solving the following
combinatorial problem (closely related to the so called
\textit{partition problem}): Given a set
$\mathcal{K}=\{k_1,\ldots,k_N\}$ of $N$ --possibly equal-- natural numbers,
how many different partitions of
$\mathcal{K}$ into two disjoint sets $\mathcal{K}_1$ and
$\mathcal{K}_2$ such that $\sum_{k\in \mathcal{K}_1}k=\sum_{k\in
\mathcal{K}_2}k$ do exist? The answer to this question can be found
in the literature (see, for example, \cite{DeRaedt} and references
therein) and is the following
\begin{eqnarray}
\label{DLcounting}
\frac{2^N}{M}\sum_{s=0}^{M-1}\prod_{I=1}^{N}\cos(2\pi s k_I/M)\ ,
\end{eqnarray}
where $M=1+\sum_{I=1}^N k_I$. This expression can be seen to be zero
if there are no solutions to the projection constraint.

Let us consider now the GM proposal.  The problem
is equivalent in this case to counting the number of irreducible representations,
taking into account multiplicities, that appear in the tensor
product $\bigotimes_{I=1}^N[j_I]$, where $[j_I]=[k_I/2]$ denotes
the irreducible representation of $SU(2)$ corresponding to spin
$j_I$. In order to solve this problem we rely on techniques
developed in the context of conformal field theories
\cite{difrancesco} (see also \cite{m}) and in the spectral theory of Toeplitz
matrices \cite{bottcher}. The starting point is to write the tensor
product of two $SU(2)$ representations in the form
$$
\left[\frac{k_1}{2}\right]\otimes\left[\frac{k_2}{2}\right]=\bigoplus_{k_3=0}^\infty\mathcal{N}^{k_3}_{k_1k_2}\left[\frac{k_3}{2}\right]\ ,
$$
where the integers $\mathcal{N}^{k_3}_{k_1k_2}$, called
\textit{fusion numbers} \cite{difrancesco},  tell us the number of
times that the representation labeled by $k_3/2$ appears in the
tensor product of $[k_1/2]$ and $[k_2/2]$. For each $k\in
\mathbb{N}\cup\{0\}$,
we introduce now the infinity \textit{fusion matrices}
$(C_k)_{k_1k_2}:=\mathcal{N}^{k_2}_{k_1k}$, where $k_1$, $k_2\in
\mathbb{N}\cup\{0\}$. These can be shown to satisfy the following
recursion relation
\begin{equation}
C_{k+2}=X C_{k+1}-C_k,\quad k=0,1,\ldots\label{recurrencia}
\end{equation}
where we have introduced the notation $X:=C_1$. Explicitly
$X_{k_1k_2}=\delta_{k_1,k_2-1}+\delta_{k_1,k_2+1}$, which shows that
$X$ is a Toeplitz matrix \cite{bottcher}. The solution to
(\ref{recurrencia}), with initial conditions $C_0=I$ and $C_1=X$,
can be written as
$$
C_k=U_k(X/2),\quad k=0,1,\ldots
$$
in terms of the Chebyshev polynomials of the second kind $U_k$. The
tensor product of an arbitrary number of representations can be
decomposed as a direct sum of irreducible representations by
multiplying the fusion matrices introduced above. By proceeding in
this way we get
$$
\left[\frac{k_1}{2}\right]\otimes \left[\frac{k_2}{2}\right]
\otimes\cdots \otimes
\left[\frac{k_N}{2}\right]=\bigoplus_{k=0}^\infty(C_{k_2}C_{k_3}\cdots
C_{k_N})_{k_1 k}\left[\frac{k}{2}\right].
$$
Notice that the product of matrices appearing in the previous
formula is, in fact, a polynomial in $X$.  The
total number of representations, that gives the solution to the
combinatorial problem at hand, is simply given by
\begin{eqnarray}
\sum_{k=0}^\infty(C_{k_2}C_{k_3}\cdots C_{k_N})_{k_1 k}.
\label{prodC}
\end{eqnarray}
This is just the sum of the (finite number of non zero) elements in
the $k_1$ row of the matrix $C_{k_2}C_{k_3}\cdots C_{k_N}$. A useful
integral representation for this sum can be obtained by introducing a
resolution of the identity for $X$ as in \cite{bottcher} and the
well-known identity $U_n(\cos\theta)=\sin[(n+1)\theta]/\sin\theta$.
In fact, equation (\ref{prodC}) can be equivalently written as
\begin{equation}
\hspace{-0cm}\frac{2}{\pi}\!\int_0^\pi \!\!\!\! \mathrm{d}\theta
\cos\frac{\theta}{2}\left[\cos\frac{\theta}{2}-\cos\big(K\!\!+\!\frac{3}{2}\big)\theta\right]\!\!
\prod_{I=1}^N\!\frac{\sin(k_I+1)\theta}{\sin\theta}
\!\!\ ,
\label{integral}
\end{equation}
where $K=k_1+\cdots+k_N$. This is related to the well known Verlinde
formula for $SU(2)$ \cite{difrancesco}.

The procedure to calculate the black hole spectrum described above
can be efficiently implemented in a computer, for
instance using \textit{Mathematica}. This allows us to analyze in detail the different factors that shape the degeneracy spectrum. First of all, the fact that the
diophantine equations are decoupled allows us to obtain the configurations compatible with a given value of area $A=\sum_{i=1}^r{q_i\sqrt{p_i}}$ as the cartesian product of the sets of
solutions to the diophantine equations for each $p_i$.
Let us then begin by analyzing the results for area values of the form $A=q\sqrt{p}$, with $q\in\mathbb{N}$ and $\sqrt{p}$ a fixed SRSFN.  What we see in this case is that the r-degeneracy
--coming from the reordering of puncture labels-- will be maximized by those
configurations having both a large number of different values of $k$ and
a large number of punctures. For a fixed area value these two factors compete with each
other because higher values of $k$ imply a lower number of punctures. On
the other hand the m-degeneracy shows an exponential growth with area (both in the DLM and GM countings).
When the two sources of degeneracy are taken into account --in the present case involving a single SRSFN--
the total degeneracy can be seen to be dominated by the m-degeneracy. The reason for this dominance of the m-degeneracy is that the number of different (small) values of $k$ available within the set of solutions to the Pell equation for a given $p$ is
limited, and hence only a few possibilities of reordering exist.

\begin{figure}[htbp]
\begin{center}
\includegraphics[width=0.48\textwidth]{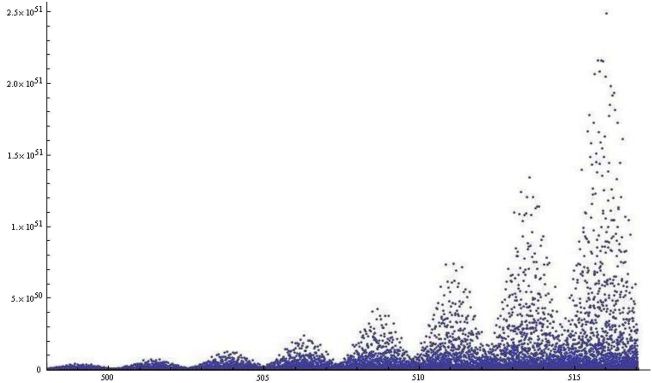}
\caption{Plot of the black hole degeneracy as a function of the
area (expressed in units of $\ell_P^2$ to facilitate the comparison with the results obtained in \cite{val1}). }\label{Fig2}
\end{center}
\end{figure}

This situation is expected to change drastically when we consider areas $A=\sum_{i=1}^r q_i
\sqrt{p_i}$, with $r>1$, built as
linear combinations of different SRSFN's. In this case it is possible to obtain configurations
with a large number of different small values of $k$ (associated to
different SRSFN's). The effect of considering linear combinations involving several SRSFN's produces a very distinctive feature when the r-degeneracy is plotted as a function of area, namely, it creates a ``band structure'' where high values of degeneracy alternate with much lower ones. Furthermore, maxima and minima are evenly spaced.
When this behavior is considered together with the m-degeneracy we obtain the
regular pattern shown in Figure \ref{Fig2}.

Several remarks are in order now. First, we want to point out that the
result obtained from the explicit computational analysis carried out in
\cite{val1} (by using the GM counting) is exactly recovered with the new approach. The fact that the same result is obtained from two completely independent procedures (a brute force approach and the algorithm proposed here) provides strong evidence for the reliability of both computations.
Second, the structure of the degeneracy spectrum obtained by using the DLM and GM countings is basically equal. They differ only in the absolute values of the degeneracy whereas the band structure (including the position and spacing of the bands) is the same.
This can be understood in our framework because the terms accounting for the r-degeneracy, responsible for
this effect, coincide for both counting procedures. This
justifies the appearance of the constant $\chi$ obtained in
\cite{val1, val2}. Third, once we understand how the
r-degeneracy works, we see that the area values for which the degeneracy is large are those that can be written as linear combinations of the SRSFN's originating from small solutions $k$
to the corresponding Pell equation. Thus, considering these linear
combinations will suffice to account for the band structure. The remaining area values give rise only to very low degeneracies.

Summarizing, we have been able to find a number-theoretic/combinatorial way to tackle the problem of calculating the degeneracy spectrum of spherical black holes in LQG. Our procedure has several advantages over previous approaches. First, we have been able to characterize the area spectrum in a proper way, giving an algorithm to explicitly find
every single spin configuration contributing to each value of the area spectrum. In particular, the degeneracies of the area eigenvalues can be obtained. This has allowed us to reproduce and understand the band structure already observed in \cite{val1} for the black hole degeneracy spectrum  in a much more efficient way. We not only recover previous results obtained by using a \emph{brute force} algorithm, but easily extend them to area values significantly larger than those reached in \cite{val1} (see Figure \ref{Fig2}).
Moreover, with our methods it is possible to compute the configurations and
degeneracy even for much larger values of area. As a token we give the degeneracy for an area of
$8320\sqrt{2}+14400\sqrt{3}+2240\sqrt{6}+4640\sqrt{15}+1120\sqrt{35}$, which is
$3.46437296507975\cdots\times10^{24420}$.
Finally, the concrete procedures and explicit formulas given in the letter offer a good starting point to study  the asymptotic behavior of the entropy as a function of the area of a black hole. This could help us
investigate whether the effective entropy quantization discussed here is present in macroscopic black holes.

\begin{acknowledgments}
We want to thank J. Baez, I. Garay, H. Sahlmann, and J. Salas for their comments.
This work is supported by the Spanish MEC grants FIS2005-05736-C03-02, FIS2005-05736-C03-03, ESP2005-07714-C03-01, and FIS2005-02761. IA and JD acknowledge financial support provided by the Spanish Ministry of Science and Education under the FPU program.

\end{acknowledgments}

\end{document}